\documentclass[preprint2]{aastex}
\usepackage{aastexug}

\newcommand{\rhovec}{\mbox{\boldmath $\rho$}}

\def\dslash{\mathbin{/\mkern-4mu/}}

\newcommand{\uvec}{\mbox{\boldmath $u$}}

\newcommand{\xvec}{\mbox{\boldmath $x$}}
\newcommand{\yvec}{\mbox{\boldmath $y$}}
\newcommand{\dvec}{\mbox{\boldmath $d$}}

\newcommand{\te}{t_{\rm E}}
\newcommand{\retilde}{\tilde{r}_{\rm E}}
\newcommand{\thetae}{\theta_{\rm E}}

\def\eqalign#1{\null\,\vcenter{\openup\jot
        \ialign{\strut\hfil$\displaystyle{##}$&$
        \displaystyle{{}##}$\hfil \crcr#1\crcr}}\,}

\begin{document}

\title{Gravitational Microlensing: A Tool for Detecting and 
       Characterizing Free-Floating Planets}

\author{Cheongho Han, Sun-Ju Chung, Doeon Kim}
\affil{Department of Physics, Institute for Basic Science Research,
       Chungbuk National University, Chongju 361-763, Korea;
       cheongho,sjchung,dekim@astroph.chungbuk.ac.kr}

\author{Byeong-Gon Park}
\affil{Bohyeonsan Optical Astronomy Observatory, Korea Astronomy
       Observatory, Youngchon 770-820, Korea;
       bgpark@boao.re.kr}

\author{Yoon-Hyun Ryu}
\affil{Department of Astronomy \& Atmospheric Science, Kyungpook National 
       University, Daegu 702-701, Korea;
       yhryu@rose0.knu.ac.kr}

\author{Sangjun Kang}
\affil{School of Liberal Arts, Semyung University, Jechon 390-711, Korea;
       sjkang@semyung.ac.kr}

\author{Dong Wook Lee}
\affil{Astrophysical Research Center for the Structure and Evolution of 
       the Cosmos (ARCSEC''), Sejong University, Seoul 143-747, Korea;
       dwlee@arcsec.sejong.ac.kr}

\begin{abstract}

Various methods have been proposed to search for extrasolar planets.
Compared to the other methods, microlensing has unique applicabilities 
to the detections of Earth-mass and free-floating planets.  However, 
the microlensing method is seriously flawed by the fact that the 
masses of the detected planets cannot be uniquely determined.  
Recently, Gould, Gaudi, \& Han introduced an observational setup 
that enables one to resolve the mass degeneracy of the Earth-mass 
planets.  The setup requires a modest adjustment to the orbit of an 
already proposed microlensing planet-finder satellite combined with 
ground-based observations.  In this paper, we show that a similar 
observational setup can also be used for the mass determinations of 
free-floating planets with masses ranging from 
$\sim 0.1\ M_{\rm J}$ to several Jupiter masses.  If the proposed 
observational setup is realized, the future lensing surveys will 
play important roles in the studies of Earth-mass and free-floating 
planets, which are the populations of planets that have not been 
previously probed.

\end{abstract}


\section{Introduction}

High-precision radial velocity measurements of nearby stars resulted 
in the detections of more than 100 planetary systems 
(http:$\dslash$cfa-www.harvard.edu/planets).  Besides the radial 
velocity method, planets can also be detected and characterized by 
using various other methods.  These methods include the pulsar timing 
analysis, direct imaging, accurate measurement of astrometric 
displacement, planetary transit, and microlensing (see the review of 
Perryman 2000).

However, all these methods except the microlensing method can be 
used to detect only planets which are gravitationally bound to their 
parent stars.  Theories about the planet formation indicate that 
planets often do not stay in the same orbit where they were formed.
Dynamical interactions with either a massive planetesimal disk
\citep{murray1998, trilling2002} or the protoplanetary nebular gas 
\citep{ward1997} can result in inward migration of planets.  On the 
other hand, gravitational interactions with other planet members 
\citep{levison1998} or neighboring stars \citep{sigurdsson1992} can 
result in the disruption of the planetary system and thus ejection 
of planets.  The latter process could lead to a population of 
free-floating planets.

Microlensing can be occurred by a low-mass object and thus free-floating 
planets can be, in principle, searched for by detecting short-duration
lensing-induced light variations of background stars caused by the 
planets.  However, this method has not been used due to technical 
difficulties. These difficulties arise because of the rarity of 
lensing events combined with the short durations of planetary events.
At any given time, only a few out of $10^6$ Galactic bulge stars 
undergo microlensing with their fluxes magnified enough for detection
and a small fraction of those microlensed stars 
are expected to be lensed by planetary masses.
In addition, events produced by planetary lenses last only a short 
duration of $\sim 1$ day even for an event caused by a giant planet 
(with a mass $M\sim 10^{-3} \ M_\odot$) and the duration decreases in 
proportional to the square root of the lens mass.  Therefore, 
implementing the microlensing method to search for free-floating 
planets requires a massive survey for a large number of stars 
with a very high monitoring frequency.  Several groups have been 
previously carried out and are currently performing lensing surveys 
\citep{alcock1993, aubourg1993, udalski1993, alard1997, bond2002},
but the typical monitoring frequency of these experiments is at most 
several times per day, which is far too low to detect events caused 
free-floating planets.  To increase the monitoring frequency, the 
experiments are employing early warning systems \citep{alcock1996, 
afonso2001, bond2001, udalski1994} to issue alerts of ongoing events 
detected in the early stage of lensing magnification and follow-up 
observation programs \citep{alcock1997, rhie1999, albrow1998, yoo2003} 
to intensively monitor the alerted events hoping to detect short-duration 
planetary signals on the top of the lensing light curve produced by 
the parent star.  However, planets detectable from this type of strategy 
are also bound planets like the ones detectable from other planet search 
methods.

However, the situation will greatly change with the advent of future 
lensing surveys with very high monitoring frequencies.  Actually, 
several such experiments in space and on the ground were proposed.  
The space-based GEST mission, proposed to NASA by \citet{bennett2002a}, 
is designed to continuously monitor $\sim 10^8$ Galactic bulge 
main-sequence stars with a frequency of several times per hour by 
using a 1--2 m aperture space telescope.  Studies about the scientific 
outcome from a ground-based high-frequency lensing experiment using four 
2 m-class telescopes located at good seeing sites around the world and 
equipped with large format CCD cameras are in progress \citep{gould2003}.
From detailed simulations of Galactic bulge events detectable from the 
space observation by using the GEST, \citet{bennett2002a} estimated that
the total number of detections of events caused by free-floating planets 
during the life time of the GEST mission ($\sim 3.7$ yrs) would be 
$\sim ({\cal O)}10$ for Earth-mass planets and $\sim ({\cal O})10^2$ 
for giant planets under the assumption that one planet was ejected for 
each star.

However, the microlensing method is seriously flawed by the fact that
the detected planet's mass cannot be uniquely determined.  This defect 
arises because among the three lensing observables related to the 
physical parameters of the lens, the Einstein time scale $\te$, the 
angular Einstein ring radius $\thetae$, and the projected Einstein 
radius onto the plane of the observer $\retilde$, only $\te$ is 
routinely measured from the lensing light curve.  These three 
observables are related to the mass of the lens ($M$), relative 
source-lens parallax ($\pi_{rel}= {\rm AU}/[D_l^{-1}-D_s^{-1}]$, 
where $D_l$ and $D_s$ are the distances to the lens and source, 
respectively), and proper motion ($\mu_{rel}$) by 
\citep{gould2000}
\begin{equation}
\te = {\thetae\over \mu_{rel}},\ \ \ 
\thetae=\sqrt{4GM\pi_{rel} \over c^2\ {\rm AU}},\ \ \ 
\retilde = \sqrt{4GM\ {\rm AU} \over c^2 \pi_{rel}}.
\label{eq1}
\end{equation}
Due to the fact that $\te$ results from the combination of $M$, 
$\pi_{rel}$, and $\mu_{rel}$, short durations of events do not 
guarantee the events are caused by planets because they can be 
produced by low-mass stars or brown dwarfs with exceptionally high 
proper motions or very small Einstein ring radii.  As a result, 
even if a large number of short-duration events were detected by 
future lensing surveys, they could not be definitely identified 
as planetary events.  To isolate events caused by free-floating 
planets and determine their masses, it is required to measure the 
other observables of $\thetae$ and $\retilde$ (see \S\ 2 for the 
detailed description about the methods to measure these quantities).  
Once $\thetae$ and $\retilde$ are measured, the lens mass is uniquely 
determined by
\begin{equation}
M = {c^2\over 4G} \retilde \thetae.
\label{eq2}
\end{equation}

Recently, \citet{gould2003} pointed out that the masses of 
{\it bound} Earth-mass planets could be measured by combining a 
GEST-type satellite with ground-based observations.  They showed 
that the proper motions of these events can be measured by analyzing 
the planet-induced perturbations in lensing light curves observed 
from space.  They also showed that if the satellite is placed in 
an L2 orbit, the Earth-satellite baseline is sufficient enough to 
measure lens parallaxes.  In this paper, we show that a similar 
observational setup proposed by \citet{gould2003} can also be used 
for the mass determinations of free-floating 
planets, which are the population of planets that have not been 
previously probed.

The paper is organized as follows.  In \S\ 2, we describe the methods 
of $\retilde$ and $\thetae$ measurements and explain the reasons for 
the difficulties in measuring these quantities for events caused by 
stellar mass lenses.  In \S\ 3, we illustrate the relative easiness of 
$\retilde$ and $\thetae$ measurements for events caused by low-mass 
lenses such as free-floating planets.  In \S\ 4, we estimate the mass 
range over which the proposed mass determination method is sensitive 
by estimating the uncertainties of the recovered lens masses under the 
observational setup proposed by \citet{gould2003}.  In \S\ 5, we 
briefly discuss several problems of the proposed observational setup.  
We conclude in \S\ 6.

\section{Mass Determinations of Gravitational Lenses}

A microlensing event occurs when a compact massive object (lens) 
approaches very close to the observer's line of sight toward a 
background star (source).  Due to the lensing-induced magnification 
and the varying lens-source separation caused by the relative 
proper motion, the source star flux changes with time.  For an 
event involved with a point source, the magnification of the 
source star flux varies with time by 
\begin{equation}
A = {u^2+2 \over u\sqrt{u^2+4}};\qquad
\uvec = \left( {t-t_0\over \te}\right)\ \hat{\xvec} 
+ u_0\ \hat{\yvec},
\label{eq3}
\end{equation}
where 
$\uvec$ represents the lens-source separation vector normalized by 
$\thetae$, $u_0$ is the closest lens-source separation in units of 
$\thetae$ (impact parameter), $t_0$ is the time at that moment 
(time of maximum magnification), and $\hat{\xvec}$ and $\hat{\yvec}$ 
are the unit vectors parallel with and normal to the direction of the 
lens-source motion, respectively.  Among the three lensing parameters 
of a point-source event ($t_0$, $u_0$ and $\te$), the first two tell 
nothing about the lens and the third is related to $M$, $\pi_{rel}$, 
and $\mu_{rel}$ in a complicated way presented in equation (\ref{eq1}).  
To fully break the lens parameter degeneracy, two additional pieces 
of information of $\thetae$ and $\retilde$ are needed.

The angular Einstein ring radius can be measured by scaling $\thetae$
against some known ``standard angular rulers'' on the source plane
(the plane perpendicular to the line of sight toward the source at the 
position of the source).  The most ubiquitous standard ruler is the 
source star itself.  If the lens passes close enough to the source, 
the source star can no longer be approximated as a point source.  
In this case, different parts on the source star surface are magnified 
by different amounts due to the difference in separation from the 
lens, and the resulting light curve deviates from that of a point 
source event; finite source effect \citep{witt1994}.  By measuring 
the deviation caused by the finite source effect, one can measure 
the relative source size normalized by the angular Einstein ring 
radius, $\rho_\star=\theta_\star/\thetae$ (normalized source radius), 
where $\theta_\star$ is the angular radius of the source star.  Since 
$\theta_\star$ can be estimated from the position of the source star 
on a color-magnitude diagram (CMD) relative to the center of red 
clump giants, whose dereddened and calibrated position on the CMD is 
well known \citep{an2002, vanbelle1999}, the angular Einstein ring 
radius is determined by $\thetae=\theta_\star/\rho_\star$ 
\citep{gould1994a, nemiroff1994}.

The projected Einstein ring radius can be measured by scaling $\retilde$ 
against some known ``standard physical rulers'' on the plane of the 
observer (the plane perpendicular to the line of sight toward the 
source at the position of the observer).  Such a standard ruler is 
provided by observing an event at two different locations.  If 
an event is seen by two observers located at different locations,
the source star position with respect to the lens is different.
The amount of the normalized difference between the source 
positions is 
\begin{equation}
\Delta \uvec 
= {\tilde{\dvec}\over \retilde},
\label{eq4}
\end{equation}
where $\tilde{\dvec}$ is the separation vector between the two 
observers projected onto the observer plane.  Then, if $\tilde{d}$ is 
large enough, the light curves of the event seen from the two 
observers will exhibit noticeable differences; parallax effect 
\citep{refsdal1966, 
gould1992}.  By measuring the difference in impact parameters, 
$\Delta u_0$, and the difference in times of maximum magnification, 
$\Delta t_0$, between the light curves as seen from the two observers, 
one can measure $\Delta\uvec = (\Delta t_0/\te,\Delta u_0)$, and so 
$\retilde$ (with the known separation vector $\tilde{\dvec}$).

However, determining the lens masses by using the mentioned standard 
rulers for events caused by stellar mass lenses was not easy.  This is 
because the Einstein ring radii of these events are much larger than 
the sizes of the rulers.
For Galactic bulge events, the sizes of 
$\thetae$ and $\retilde$ are
\begin{equation}
\thetae = 1.26\ {\rm mas}\ 
\left( {D_s\over D_l}-1 \right)^{1/2}
\left( {D_s\over 8\ {\rm kpc}} \right)^{-1/2}
\left( {M\over M_\odot} \right)^{1/2},
\label{eq5}
\end{equation}
and 
\begin{equation}
\retilde = 8.2\ {\rm AU}\ 
\left( {D_s\over D_l}-1 \right)^{-1/2}
\left( {D_s\over 8\ {\rm kpc}} \right)^{1/2}
\left( {M\over M_\odot} \right)^{1/2}.
\label{eq6}
\end{equation}
Noticeable light curve deviations caused by the finite source effect 
are produced only when the lens-source separation is equivalent to 
or less than the source radius (see Figure \ref{f2}).
However, the normalized source radius of a typical bulge event caused 
by a stellar mass lens with $M\sim 0.3\ M_\odot$ located at $D_l=
6\ {\rm kpc}$ is merely $\rho_\star \sim 0.0018$ for a main-sequence 
star (with a physical radius of $R_\star\sim 1\ R_\odot$ and an angular 
size of $\theta_\star\sim 0.6\ \mu$arcsec at $D_s\sim 8\ {\rm kpc}$) and 
$\rho_\star \sim 0.02$ even for a clump giant star (with $R_\star\sim 
13\ R_\odot$ and $\theta_\star\sim 7.6\ \mu$arcsec).  
This implies that close lens-source approach enough 
to produce noticeable light curve deviations caused by the finite source 
effect occurs very rarely and thus one can measure $\thetae$ only for 
a very small fraction of events.  In addition, the typical projected 
Einstein ring radius for these events is $\retilde \sim 7.6\ {\rm AU}$, 
and thus measurement of $\retilde$ is impossible with an Earth-scale 
baseline.\footnote{For an event with a very long time scale 
($\te\gtrsim 100$ days), it is possible to measure $\retilde$ because 
the location of the observer changes considerably during the event 
because of the orbital motion of the Earth around the Sun.} One may 
think an appropriate baseline can be provided by a satellite with a 
heliocentric orbit.  However, the problem of this solution is that 
transmission of a huge amount of data from the satellite is difficult 
with the current technology.  In addition, even if the technical 
problem is resolved and events are observed from the Earth and the 
satellite, still $\retilde$ cannot be uniquely determined because 
of the inherent degeneracy in $\Delta\uvec$.  This degeneracy arises 
because for a given pair of light curves observed from the two 
different locations there can be two possible values of $\Delta u$ 
depending on whether the source star trajectories as seen from the 
Earth and the satellite are on the same or opposite sides with 
respect to the lens, causing a two-fold degeneracy in the determined 
value of $\retilde$ \citep{gould1994b}.  To lift this degeneracy, 
one needs a second satellite.  As a result, $\theta_{\rm E}$ and 
$\retilde$ have been measured for only a handful number of events 
each out of the more than 1000 events detected so far 
[\citet{alcock1997,alcock2000, albrow1999, albrow2000, albrow2001, 
afonso2000, yoo2003} for $\theta_{\rm E}$ measurements and 
\citet{alcock1995, mao1999, soszynski2001, bond2001, mao2002, 
bennett2002b} for $\retilde$ measurements], and for only a single 
event both $\thetae$ and $\retilde$ have been measured 
\citep{an2002}.\footnote{We note that most of the events for which 
$\thetae$ are measured are caustic-crossing binary lens events, 
which comprise only a small fraction of the total number of events.  
We also note that all of the events with known parallaxes are very 
long time-scale events, which comprise also a small fraction of 
the total number of events.}

\section{Events Caused by Free-Floating Planets}

Unlike events caused by stellar mass lenses, it is possible to 
measure both $\retilde$ and $\thetae$ for a substantial fraction 
of events caused by free-floating planets.  This is because the 
Einstein ring radii of these events are much smaller than those 
of the events produced by stellar mass lenses.  Then, the sizes 
of the angular and physical standard rulers on the source and 
observer planes compared to $\thetae$ and $\retilde$, respectively, 
are no longer negligible, and thus the events are more susceptible 
to both the finite-source and parallax effects.

\subsection{Parallax Effect}

If an event caused by a free-floating planet is observed simultaneously 
from the Earth and a satellite located at a distance of $d\sim 0.01$ AU, 
corresponding to the distance to a satellite located in the L2 orbit, 
the expected displacement between the locations of the source caused 
by the parallax effect for a typical bulge event (with $D_s\sim 8$ kpc 
and $D_s\sim 6$ kpc) would be
\begin{equation}
\Delta u = {d \cos\psi \over \retilde} \sim 
0.02 \cos \psi \left( {M \over M_{\rm J}} \right)^{-1/2},
\label{eq7}
\end{equation}
where $\cos \psi$ is the projection factor.  Although this shift 
is significantly larger than the shift expected from the event 
caused by a stellar mass lens, it is still small compared to 
$\thetae$.  Then, one may question whether the light curves observed 
from the Earth and the satellite can exhibit noticeable differences 
with such a small displacement.

For high magnification events, however, the light curve difference 
induced by the parallax effect can be large even with a small amount 
of $\Delta u$.  The reason can be analytically explained as follows.  
Differentiating the expression of the magnification in equation 
(\ref{eq3}) with respect to $u$ yields
\begin{equation}
{dA\over du} = -{8 \over u^2(u^2+4)^{3/2}}.
\label{eq8}
\end{equation}
Then the fractional difference between the light curves seen from the 
Earth and the satellite is computed by
\begin{equation}
\epsilon_\pi = {A_{sat}-A_\oplus\over A_\oplus} 
\sim 
{\Delta u\over A_\oplus}\left\vert {dA_\oplus\over du}\right\vert
= {8\Delta u \over u(u^2+2)(u^2+4)},
\label{eq9}
\end{equation}
where $A_\oplus$ and $A_{sat}$ represent the magnifications measured 
from the Earth and the satellite, respectively.  For a high magnification 
event ($u\rightarrow 0$), equation (\ref{eq9}) is simplified into the form 
\begin{equation}
\epsilon_\pi \sim {\Delta u\over u}.
\label{eq10}
\end{equation}
From equation (\ref{eq10}), one finds that even with a small amount of
$\Delta u$, the deviation can be large for high magnification events.  
For example, the fractional difference between the light curves caused 
by the source star's positional displacement of $\Delta u=0.02$ is 
$\epsilon_\pi \gtrsim 5\%$ for events with $u_0 \lesssim 0.4$, implying 
that the difference can be measured for a significant fraction of 
events caused by free-floating planets.

%Figure 1
\begin{figure}[t]
\epsscale{1.1}
\vskip-0.7cm
\centerline{\plotone{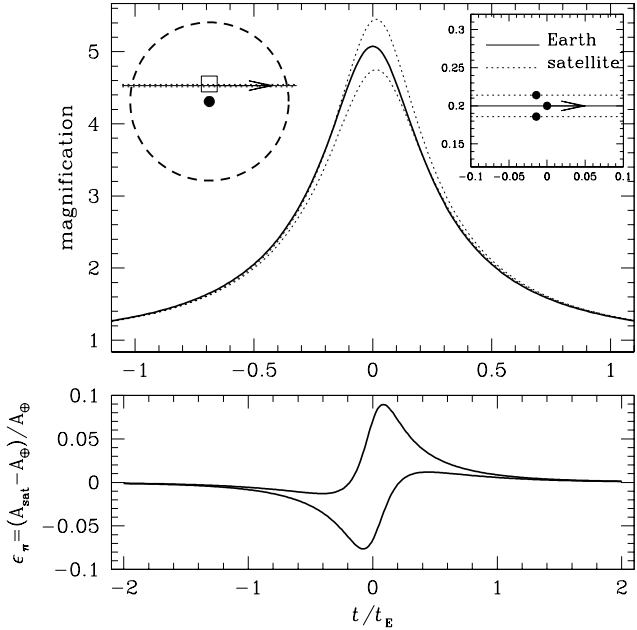}}
\vskip-0.6cm
\caption{
Upper panel: 
Light curves of an example planetary event observed from the Earth 
(solid curve) and a satellite (dotted curves).  The inset on the  
left corner shows the geometry of the lens system, where the dashed 
circle represents the Einstein ring and the solid and dotted straight 
lines represent the source star trajectories as seen from the Earth 
and the satellite, respectively, with respect to the lens (marked by 
a small filled dot at the center of the Einstein ring).  The inset 
on the right corner shows the blowup of the region enclosed by a 
square on the left-side inset.  The three dots represent the source 
locations at the moment when the magnification of the event seen 
from the Earth is maximum.
Lower panel: 
Fractional difference between the light curves.
}
\label{f1}
\end{figure}

The feasibility of parallax measurements for high magnification events 
is illustrated in Figure \ref{f1}, where we present light curves of an 
example planetary event observed from the Earth (solid curve) and a 
satellite (dotted curves).  The event as seen from the Earth has an 
impact parameter of $u_0=0.2$ and the source position as seen from 
the satellite is slightly displaced by $\Delta u_0=0.02/\sqrt{2}$ and 
$\Delta t_0/\te=0.02/\sqrt{2}$ (and thus $\Delta u=\sqrt{(\Delta t_0/\te)^2
+\Delta u_0^2}=0.02$).  The inset on the left corner of the upper panel 
shows the geometry of the lens system, where the dashed circle represents 
the Einstein ring and the solid and dotted straight lines represent the 
source star trajectories as seen from the Earth and the satellite, 
respectively, with respect to the lens (marked by a small filled dot at 
the center of the Einstein ring).  The inset on the right corner shows 
the blowup of the region enclosed by a square in the left-side inset.  
The three dots represent the source locations at the moment when the 
magnification of the event seen from the Earth is maximum, i.e.\ $t=t_0$.  
The lower panel shows the fractional deviation $\epsilon_\pi$.  One 
finds that even with a very small displacement of the source trajectory, 
the deviation is as large as $\epsilon_\pi \sim 10\%$ as predicted by 
equation (\ref{eq10}).

We note that the parallax of the event caused by a free-floating planet 
does not suffer from the degeneracy mentioned in \S\ 2.  This is because 
the source trajectory seen from the Earth and the satellite are nearly 
always located on the same side with respect to the lens due to the 
small value of $\Delta u$ expected for typical planetary events.

%Figure 2
\begin{figure}[t]
\vskip-0.7cm
\centerline{\plotone{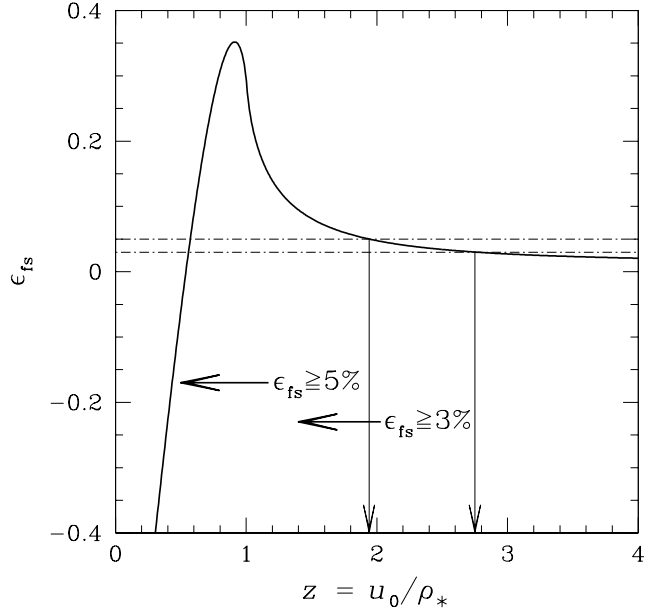}}
\vskip-0.9cm
\caption{
Fractional deviation of the lensing magnification caused by the 
finite source effect as a function of $z=u_0/\rho_\star$, where $u_0$ 
and $\rho_\star$ represent the normalized impact parameter of the 
lens-source encounter and the normalized source radius.  Then, $z<1$ 
($u_0<\rho_\star$) implies that the lens transverse the source star 
surface.
}
\label{f2}
\end{figure}

\subsection{Finite Source Size Effect}

High-magnification planetary events are susceptible not only to 
the parallax effect but also to the finite source effect.  The 
light curve affected by the finite source effect is given by the 
intensity-weighted magnification averaged over the source flux.  
Under the assumption of a uniform source brightness, the 
magnification of an event affected by the finite source effect 
is expressed by
\begin{equation}
A_{fs}
= {1\over \pi\rho^2_\star}
\int_0^{2\pi} \hskip-6pt d\theta 
\int_0^{\rho_\star} \hskip-6pt d\rho\ \rho
A\left[\sqrt{(u_c+\rho\cos\theta)^2+(\rho\sin\theta)^2}\right],
\label{eq11}
\end{equation}
where $u_c$ represents the normalized separation between the lens and 
the center of the source star, and $\rhovec$ and $\theta$ represent 
the position vector and the position angle of a point on the source 
star surface with respect to the source star's center, respectively.  
In the case of a high magnification event, for which the finite source 
effect is important, the magnification is approximated by  
\begin{equation}
A \sim {1\over u}.
\label{eq12}
\end{equation}
Using equations (\ref{eq11}) and (\ref{eq12}), one finds the expression 
for the ratio between the magnifications with and without the finite 
source effect, i.e.
\begin{equation}
{A_{fs}\over A} \sim B(z) =
{z^2\over \pi}
\int_0^{2\pi} \hskip-6pt d\theta \ 
\int_0^{1/z} \hskip-6pt d\zeta 
{\zeta\over \sqrt{\zeta^2 + 2\zeta\cos\theta +1}},
\label{eq13}
\end{equation} 
where $z=u/\rho_\star$ \citep{gould1994a}.
Then, the fractional deviation caused by the finite source effect 
is expressed in terms of the function $B(z)$ as
\begin{equation}
\epsilon_{fs}={A_{fs}-A\over A} \sim B(z)-1.
\label{eq14}
\end{equation}

In Figure \ref{f2}, we present the fractional deviation caused by the 
finite source effect as a function of $z$.  From the curve, one finds 
that if the lens approaches the source star closer than $\sim 1.9$ 
and $\sim 2.7$ times of the source star radius, the deviations become 
$\epsilon_{fs}\gtrsim 5\%$ and $\gtrsim 3\%$, respectively.  The 
normalized angular radius of a Galactic bulge main-sequence source 
star (with a physical size of $R_\star\sim 1\ R_\odot$ and 
an angular size of 0.6 $\mu$arcsec at $D_s\sim 8\ {\rm kpc}$), 
which is the main monitoring target of the GEST mission, is 
$\rho_\star\sim 0.032 (M/M_{\rm J})^{-1/2}$.  Then, with a detection 
condition of $\epsilon_{fs}\geq 3\%$, the deviation can be detected 
for events with $u_0 \lesssim 2.7\times 0.032 (M/M_{\rm J})^{-1/2}
\sim 0.086 (M/M_{\rm J})^{-1/2}$.  Note that while one needs both 
light curves obtained from the Earth and the satellite for $\retilde$ 
measurement, $\thetae$ can be measured from a single light curve 
obtained from space observation with exquisite photometry.

\section{Sensitivity: Quantitative Analysis}
As the mass of the lens decreases, the resulting event becomes more
susceptible to the parallax effect for a given baseline and to the 
finite source effect for a given source radius.  However, if the lens 
mass is too small, the detection efficiency of the deviation will 
sharply drop and the uncertainty of the determined lens mass will 
become large due to not-enough coverage of the deviation.  In this 
section, we estimate the lens mass range over which the proposed 
method is sensitive.  For this, we estimate the uncertainties of 
$\retilde$ and $\thetae$ expected under the observational setup 
proposed by \citet{gould2003}.

To determine the uncertainties of $\retilde$ and $\thetae$, we first 
estimate the uncertainties of the lensing parameters.  We estimate 
these uncertainties by computing the curvature matrix of $\chi^2$ 
surface, $c_{ij}$.  
For the case of a microlensing light curve, the curvature matrix 
is defined by 
\begin{equation}
c_{ij}=\sum_k^{N_{obs}} {\partial F_{obs,k}\over \partial p_i}
{\partial F_{obs,k}\over \partial p_j}
{1\over \sigma_k^2},
\label{eq15}
\end{equation}
where $N_{obs}$ is the number of observations, $F_{obs,k}(t)=A(t)F_s+F_b$ 
is the observed flux for each measurement, $p_i=(F_s,F_b,u_0,t_0,t_{\rm E},
\rho_\star)$ are the lensing parameters, $F_s$ and $F_b$ are the fluxes 
of the lensed source star and blended light, and $\sigma_k$ is the 
photometric uncertainty of each measurement.  We note that in addition 
to the lensing parameters of a point-source event, two additional 
parameters of $F_b$ and $\rho_\star$ are included to account for the 
blending and finite source effects.  Then, the uncertainties of the 
individual lensing parameters correspond to the diagonal components 
of the inverse curvature matrix (covariance matrix $b=c^{-1}$, where 
$\sum_k b_{ik}c_{kj}=\delta_{ij}$), i.e.
\begin{equation}
\sigma_{p_i} = \sqrt{b_{ii}}.
\label{eq16}
\end{equation}

The error analysis is done for events with a source and a lens located
at 8 kpc and 6 kpc, respectively, and varying the lens mass
and $u_0$.
The assumed relative proper motion is $\mu_{rel}=7.0$ mas/yr
(corresponding to the relative lens-source transverse speed of 
$v_{rel}=200\ {\rm km\ s}^{-1}$).
Then, the corresponding event time scale is $\te=0.96\ {\rm days}\ 
(M/M_{\rm J})^{1/2}$.
The proposed 
GEST mission will take images of Galactic bulge main-sequence stars 
with 2 min exposure and coadd 5 images to yield 10 min exposure
\citep{bennett2002a}.  Following the specification of the GEST 
mission, we assume that each combined image is obtained every 20 min 
(corresponding to a monitoring frequency of 72/day) considering the 
data readout time.  We also assume a similar monitoring frequency for 
the ground observation.  
We assume each event is observed during $-2\te \leq t_{obs} < 2\te$.
For the photometric uncertainties, we assume 
1\% and 2\% photometry for the space and ground observations, 
respectively.  We also assume that blended light comprises 0\% (no 
blending) and 30\% of the observed flux for the space and ground 
observations, respectively.  The source star is assumed to have
a solar radius.

Once $\sigma_{p_i}$ expected from the space and ground observations 
are determined, the uncertainty of the source trajectory displacement 
($\Delta u$) is computed according to error propagation analysis by
\begin{equation}
\sigma_{\Delta u} \sim  
{
\left[ 
\sigma_{\Delta t_0 / \te}^2 ({\Delta t_0 / \te})^2
+
\sigma_{\Delta u_0}^2 \Delta u_0^2
\right]^{1/2} 
\over 
\Delta u},
\label{eq17}
\end{equation}
where $\sigma_{\Delta t_0 / \te}$
and $\sigma_{\Delta u_0}$ are the uncertainties of the differences 
between the $t_0/t_{\rm E}$'s and $u_0$'s measured from the two 
light curves obtained from the space and ground observations.  To 
prevent confusion in notations, we note that ``$\sigma$'' is 
used to denote the measurement uncertainty, while ``$\Delta$'' is 
used to denote the difference in lensing parameters determined from 
the light curves obtained from the space and ground observations.  
The uncertainties $\sigma_{\Delta t_0 / \te}$ and $\sigma_{\Delta u_0}$ 
are computed from $\sigma_{p_i}$'s estimated from curvature matrix
analysis by  
\begin{equation}
\eqalign{
 & \sigma_{\Delta t_0/\te}^2 
\sim 
\left[ \left( {\sigma_{\te}\over \te} \right)_{sat}^2 +
\left( {\sigma_{\te}\over \te} \right)_\oplus^2 \right] 
\left({\Delta t_0 \over \te} \right)^2,
\cr
 & \sigma_{\Delta u_0}^2 
\sim (\sigma_{u_0})_{sat}^2 + 
(\sigma_{u_0})^2_\oplus, \cr
}
\label{eq18}
\end{equation}
where the subscripts ``{\it sat}'' and ``$\oplus$'' are used to denote 
uncertainties from space and ground measurements, respectively.
The first relation in equation (\ref{eq18}) is obtained under 
the assumption that the uncertainty of $\sigma_{\Delta t/\te}$ is 
dominated by $\sigma_{\te}$, i.e. $\sigma_{\te} \gg \sigma_{t_0}$.  
To account for the projection effect of the Earth-satellite 
separation, we assume $\tilde{d}=d/2=0.005$ AU and the displacement 
vector of the tested event is assumed to have equal $x$ and $y$ 
components, i.e.\ $\Delta t_0/\te = \Delta u_0 = \Delta u/\sqrt{2}$.  
Since the normalized source star radius can be determined from the 
light curve observed from space alone, its uncertainty is given 
simply by
\begin{equation}
\sigma_{\rho_\star} = {(\sigma_{\rho_\star})}_{sat}.
\label{eq19}
\end{equation}
Once $\sigma_{\Delta u}$ and $\sigma_{\rho_\star}$ are determined, 
the fractional uncertainties of $\retilde$, $\thetae$, and the 
resulting lens mass are obtained by
\begin{equation}
\eqalign{
{\sigma_{\retilde} \over \retilde} & \sim {\sigma_{\Delta u}
        \over \Delta u}, \cr
{\sigma_{\thetae} \over \thetae}  & \sim {\sigma_{\rho_\star} 
        \over \rho_\star}, \cr
{\sigma_M \over M}  & \sim  \left[
\left({\sigma_{\retilde} \over \retilde}\right)^2  +
\left({\sigma_{\thetae} \over \thetae}\right)^2 \right]^{1/2}.
\cr
}
\label{eq20}
\end{equation}

%Figure 3
\begin{figure}[t]
\vskip-0.2cm
\epsscale{1.46}
\centerline{\plotone{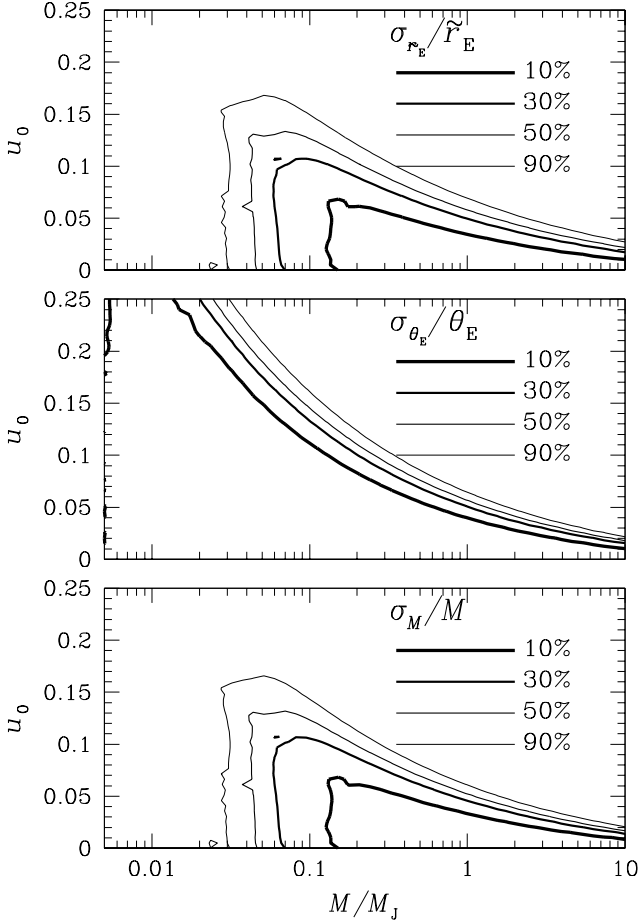}}
\vskip-0.3cm
\caption{
Contours maps of the fractional uncertainties of $\retilde$ (upper 
panel), $\thetae$ (middle panel), and the lens mass (lower panel)
in the parameters space of $(u_0, M)$.
}
\label{f3}
\end{figure}

In Figure 3, we present the estimated uncertainties for $\retilde$ 
(upper panel), $\thetae$ (middle panel), and the lens mass (lower 
panel) as contours maps in the parameters space of $(u_0, M)$.  
In each map, contours are drawn at the levels of the fractional 
uncertainties of 10\%, 30\%, 50\%, and 90\%.  From the figure, one
finds that the proposed method is most sensitive to planets with 
masses ranging from $M\sim 0.1\ M_{\rm J}$ to several Jupiter masses.  
One also finds that while $\sigma_{\thetae}$ for a given value of 
$u_0$ decreases with the decrease of the lens mass until the lens 
mass is $M\sim 0.01\ M_{\rm J}$ (equivalent to the mass of an 
Earth-mass planet), at which the coverage of the deviation becomes 
less sufficient, the rapid increase of $\sigma_{\retilde}$ occurs 
at a substantially larger mass of $M\sim 0.1\ M_{\rm J}$.  As a result, 
the uncertainty in the determined lens mass for planets with 
$M\lesssim 1.0\ M_{\rm J}$ is mostly dominated by $\sigma_{\retilde}$ 
rather than $\sigma_{\thetae}$.  The uncertainty $\sigma_{\retilde}$ 
estimated from the numerical analysis is larger than the expectation 
based on analytic analysis in \S\ 3.1.  To track down the difference, 
we did the same curvature matrix analysis to determine 
$\sigma_{\retilde}/\retilde$ but not taking account the finite 
source effect (and thus $\rho_\star$ is excluded from the list of 
fitting parameters).  From this, we find significant decrease of 
$\sigma_{\retilde}/\retilde$ with the decrease of the number of 
fitting parameters.  We find a similar trend when the blending effect 
is not taken into consideration.  We, therefore, conclude that the 
larger uncertainty $\sigma_{\retilde}/\retilde$ obtained from the 
numerical analysis compared to the simple analytic analysis is due to 
the increase of the number of fitting parameters that are additionally 
included to account for the blending and the finite source effects
(i.e.\ $F_b$ and $\rho_\star$).

%We find two 
%reasons for the dominance of $\sigma_{\retilde}$ over $\sigma_{\thetae}$ 
%for planets in this mass range.  The first reason is the relatively 
%poor photometric precision of the ground observation compared to 
%that of the space observation.  Since $\retilde$ measurement requires 
%both light curves obtained from space and ground observations, 
%$\retilde$ measurement suffer from larger uncertainties.  Another 
%important reason is the decrease of the parallax signal, i.e.\ 
%$\epsilon_\pi$, with the increasing finite source effect.  For events
%effected by severe finite source effect, the peak magnification of 
%the resulting light curve is substantially lower than that of a 
%point-source event, and thus the parallax signal is lowered.

\section{Discussion}

Considering the high sampling rate with a large format CCD camera, 
the total data rate of the space microlensing survey will be large, 
and thus the data transmission from the satellite even at the L2 
position will be still a major concern.  One way to alleviate the problem 
is putting a satellite in a highly elliptical orbit with a period of 
$\sim 1$ month as proposed by \citet{gould2003}.  Then, the satellite 
would spend majority of its time near $2a\sim 0.005$ AU and it could 
focus on data transmission during the brief perigee each month.

Another concern for the proposed satellite observation is the location 
of the Galactic bulge field in ecliptic coordinates, which are $\lambda=
266^\circ\hskip-2pt .84$ and $\beta=-5^\circ\hskip-2pt .54$.  Then, while the 
physical distance to the L2 point is $d\sim 0.01$ AU, the projected 
separation is very small ($\tilde{d}\sim 0.001$ AU) around the time 
of summer solstice, when the terrestrial observation is optimal.  
Therefore, during this time of the year, it would be impossible to 
measure the lens parallax by using the proposed method.  However, 
the bulge season lasts more than $\sim 6$ months and the survey can 
be initiated  before spring equinox and can be extended after autumnal 
equinox, at which the projected Earth-satellite separation is maximized.  
Therefore, lens parallaxes can be measured for a large fraction of the 
observation time.

\clearpage

We note that the actual uncertainty of the mass measurements 
will be larger than our estimation because we did not include
the uncertainty of the angular source size.  There is 
no empirical calibration of stellar size for main-sequence
stars because it is beyond the limit of current interferometric measurements
and the available stellar size calibration for these stars 
are entirely based on theoretical stellar atmosphere calculations.
In addition, the uncertainty of the distance to the source star
will also contribute to the uncertainty of the mass determinations 
for free-floating planets.

\section{Conclusion}
We propose an observational setup that enables one to determine 
the masses of free-floating planets via gravitational microlensing.  
The setup is similar to the one of \citet{gould2003} proposed to 
resolve the mass degeneracy of the Earth-mass planets and requires 
a modest adjustment to the orbit of an already proposed microlensing 
planet-finder satellite combined with ground-based observations.  
The proposed method is most sensitive to planets with masses 
ranging from $\sim 0.1\ M_{\rm J}$ to several Jupiter masses.  
If realized, the proposed high-frequency lensing surveys will be 
able to play important roles in characterizing Galactic free-floating 
planets that have not been probed before.

\acknowledgments
We would like to thank B.\ S. Gaudi and A. Gould for making helpful 
comments.  This work was supported by the Astrophysical Research Center 
for the Structure and Evolution of the Cosmos (ARCSEC") of Korea Science 
\& Engineering Foundation (KOSEF) through Science Research Program (SRC) 
program.

\end{document}